\documentclass[
prl,twocolumn,showpacs,preprintnumbers,amsmath,amssymb]{revtex4}

\usepackage{graphicx}
\usepackage{amssymb}
\usepackage{amsmath}
\usepackage{dsfont}
\usepackage{bm}
\usepackage{mathrsfs}
\usepackage{times}

\begin{document}
\title{Optical angular momentum in dispersive media}

\author{T.\ G.\ Philbin}
\email{t.g.philbin@exeter.ac.uk}
\affiliation{Physics and Astronomy Department, University of Exeter,
Stocker Road, Exeter EX4 4QL, United Kingdom}

\author{O.\ Allanson}
\affiliation{School of Physics and Astronomy, University of St Andrews,
North Haugh, St Andrews, Fife KY16 9SS, United Kingdom}

\begin{abstract}
The angular momentum density and flux of light in a dispersive, rotationally symmetric medium are derived from Noether's theorem. Optical angular momentum in a dispersive medium has no simple relation to optical linear momentum, even if the medium is homogeneous. A circularly polarized monochromatic beam in a homogeneous, dispersive medium carries a spin angular momentum of $\pm\hbar$ per energy $\hbar\omega$, as in vacuum. This result demonstrates the non-trivial interplay of dispersive contributions to optical angular momentum and energy. 
\end{abstract}

\maketitle

The discovery by Allen {\it et al.}~\cite{all92} that light beams can carry orbital angular momentum has led to intensive study of the angular momentum of light and its applications (see~\cite{yao11} for a recent review). In vacuum, the linear momentum density $\mathbf{p}=\varepsilon_0\mathbf{E}\times\mathbf{B}$ of light determines its angular momentum density $\mathbf{r}\times\mathbf{p}$. A beam with azimuthal phase dependence carries orbital angular momentum in the direction of propagation, distinct from the more familiar spin angular momentum of circular polarization ($\pm\hbar$ per photon)~\cite{yao11}. Inside a medium, however, these expressions for the linear and angular momentum densities of light are no longer correct. If the beam has frequency components for which absorption is significant then there is no conserved optical linear or angular momentum. But if the beam is confined to a frequency range where losses are negligible, then light in a rotationally symmetric medium will carry a conserved angular momentum. A frequency range with negligible losses will in general exhibit dispersion, visible light in glass being the most famous example. The problem solved in this paper is the following: what is the optical angular momentum density and flux in a dispersive, rotationally symmetric medium for beams with negligible absorption? 

Even if the beam is monochromatic, dispersion contributes to the energy-momentum and angular momentum of light. This is best known from the Brillouin expression for the time-averaged energy density of monochromatic light in a dispersive medium~\cite{jac}, which depends on the derivatives $d\varepsilon(\omega)/d\omega$ and  $d\mu(\omega)/d\omega$ of the permittivity and permeability at the frequency of the beam. The  time-averaged monochromatic momentum density also depends on $d\varepsilon(\omega)/d\omega$ and  $d\mu(\omega)/d\omega$~\cite{phi11,note}. The energy density and momentum density for beams with a finite frequency range has a much more complicated dependence on the dispersion~\cite{phi11}. Here we find the angular momentum density and flux for a finite frequency range and also specialize the result to the monochromatic case. Unlike in vacuum, the angular momentum density has no simple relation to the linear momentum density (even in a homogeneous medium where the latter obeys a conservation law). This means that the conserved angular momentum cannot be constructed from the energy-momentum tensor, although this is how optical  angular momentum in media is usually addressed~\cite{pfe07}. The conserved optical angular momentum is associated with rotational symmetry and must therefore be calculated using Noether's theorem.

Note that the problem solved here is self-contained, unambiguous, and unaffected by considerations of angular momentum transfer between light and matter. We consider only the correct expression for conserved optical angular momentum; an analysis of the transfer of optical angular momentum to matter requires ingredients not needed here~\cite{pad03,man08}. The question we address is one of the most basic that can be posed for any field theory. The solution to the problem, in combination with the result for the energy-momentum tensor, demonstrates a remarkable interplay of dispersive contributions to the conserved quantities of light.

As in~\cite{phi11}, our results are derived from Noether's theorem applied to the electromagnetic action in a dispersive medium, assuming a finite frequency range with negligible losses. In such a frequency range the dielectric functions can be fitted to an even series in frequency:
\begin{equation} \label{matseries}
\varepsilon(r,\omega)=\sum_{n=0}^\infty\varepsilon_{2n}(r)\,\omega^{2n}, \quad \kappa(r,\omega)=\sum_{n=0}^\infty \kappa_{2n}(r)\,\omega^{2n},
\end{equation}
where $\varepsilon(r,\omega)$ is the relative permittivity and the relative permeability is $\mu(r,\omega)=\kappa(r,\omega)^{-1}$. In practice the series in (\ref{matseries}) will have a finite number of terms and will represent a fit to dispersion data in the frequency range of interest, but we allow an infinite number of terms to obtain the results in their greatest generality. The expansions (\ref{matseries}) are standard in treating dispersion in frequency ranges where absorption is negligible (see e.g.~\cite{agr}). To ensure the existence of a conserved optical angular momentum, the medium is taken to be rotationally symmetric, with dielectric functions (\ref{matseries}) depending on $r=\sqrt{\mathbf{r}\cdot\mathbf{r}}$. The $\mathbf{D}$ and $\mathbf{H}$ fields in the frequency domain and in the time domain are
\begin{gather*}
\mathbf{\tilde{D}}(\mathbf{r},\omega)=\varepsilon_0\varepsilon(r,\omega)\mathbf{\tilde{E}}(\mathbf{r},\omega), \  \mathbf{D}(\mathbf{r},t)=\varepsilon_0\varepsilon(r,i\partial_t)\mathbf{E}(\mathbf{r},t),    \\
 \mathbf{\tilde{H}}(\mathbf{r},\omega)=\kappa_0\kappa(r,\omega)\mathbf{\tilde{B}}(\mathbf{r},\omega),   \ \mathbf{H}(\mathbf{r},t)=\kappa_0\kappa(r,i\partial_t)\mathbf{B}(\mathbf{r},t).
\end{gather*}
where $\kappa_0=\mu_0^{-1}$. The electromagnetic action in the medium (\ref{matseries}) is~\cite{phi11}
\begin{equation} \label{act}
\mathcal{S}=\int d^4x\frac{\kappa_0}{2}\left\{\frac{1}{c^2}\mathbf{E}\cdot[\varepsilon(r,i\partial_t)\mathbf{E}]-\mathbf{B}\cdot[\kappa(r,i\partial_t)\mathbf{B}]\right\},
\end{equation}
with the dynamical variables taken to be the scalar potential $\phi$ and vector potential $\mathbf{A}$, defined by
\begin{equation} \label{pots}
\mathbf{E}=-\nabla\phi-\partial_t\mathbf{A}, \quad \mathbf{B}=\nabla\times\mathbf{A}.
\end{equation}
Variation of  $\phi$ and $\mathbf{A}$ in (\ref{act}) gives the macroscopic Maxwell equations (with no free charges or currents) 
\begin{gather} 
\varepsilon_0\nabla\cdot[\varepsilon(r,i\partial_t)\mathbf{E}]=0,  \label{max1}  \\
\kappa_0\nabla\times[\kappa(r,i\partial_t)\mathbf{B}]=\varepsilon_0\varepsilon(r,i\partial_t)\partial_t\mathbf{E}. \label{max2}
\end{gather}
The other two Maxwell equations are identities due to (\ref{pots}).

The action (\ref{act}) is invariant under active rotations of the dynamical fields $\phi$ and $\mathbf{A}$ around the origin $r=0$, and this invariance implies the existence of a conserved quantity, optical angular momentum. For a homogeneous medium the invariance of the action holds for rotations of the dynamical fields around an arbitrary point, but this point can always be chosen as the coordinate origin. An infinitesimal rotation of the scalar field $\phi$ around the origin is given by
\begin{gather}
\phi(\mathbf{r},t)\rightarrow \phi(\mathbf{r}-\delta\mathbf{r},t),  \\
\delta r^i=\Omega^i_{\ j}r^j, \quad \Omega_{ij}=-\Omega_{ji},
\end{gather}
where $\Omega_{ij}=-\Omega_{ji}$ denote the three independent infinitesimal parameters of the rotation. (We use tensor notation throughout, with indices lowed and raised by the metric $g_{ij}$ and its inverse $g^{ij}$.) The vector potential undergoes the rotation
\begin{equation}
A^i(\mathbf{r},t)\rightarrow (\delta^i_{\ j}+\Omega^i_{\ j})A^j(\mathbf{r}-\delta\mathbf{r},t),
\end{equation}
and the infinitesimal variations of $\phi$ and $\mathbf{A}$ are thus
\begin{equation}  \label{rot}
\delta\phi=-\Omega_{ij}r^j\nabla^i\phi, \qquad \delta A_i=\Omega_{ij}A^j-\Omega_{jk}r^k\nabla^jA_i.
\end{equation}
Noether's theorem~\cite{wei} guarantees that if we let the rotation parameters depend on space and time, i.e.\ $\Omega_{ij}=\Omega_{ij}(\mathbf{r},t)$, then the change in the action (\ref{act}) under (\ref{rot}) can be written in the form
\begin{equation} \label{actvar}
\delta\mathcal{S}=-\frac{1}{2}\int d^4x\left(L^{ij}\partial_t\Omega_{ij}+M^{kij}\nabla_k\Omega_{ij}\right),
\end{equation}
where the angular momentum density $L^{ij}=-L^{ji}$ and flux $M^{kij}=-M^{kji}$ obey the conservation law
\begin{equation} \label{con}
\partial_t L^{ij}+ \nabla_k M^{kij}=0.
\end{equation}
The angular momentum density $L^{ij}$ emerges as an antisymmetric tensor; its dual is a vector and will be constructed later. Integrations by parts, in which surface terms are to be dropped, are required to achieve the form (\ref{actvar}). Use must also be made of the identities (10) and (11) in ref.~\cite{phi11}. The result is as follows, where we use square brackets to denote antisymmetrization of indices~\footnote{For example, $T_{[ij]}=(T_{ij}-T_{ji})/2$.}, $\epsilon_{ijk}$ is the completely antisymmetric Levi-Civita tensor, and $\mathcal{L}$ is the Lagrangian density (i.e.\ the integrand in the action (\ref{act})):
{\allowdisplaybreaks
\begin{align}
L_{ij}=& 2\varepsilon_0A_{[j}\varepsilon(r,i\partial_t)E_{i]}+2\varepsilon_0r_{[i}\nabla_{j]}A^k\varepsilon(r,i\partial_t)E_k   \nonumber \\
+r_{[i} & \left[\varepsilon_0\sum_{n=1}^\infty\sum_{m=1}^{2n}(-1)^{n+m}\varepsilon_{2n}(r)\partial_t^{m-1}E^k\partial_t^{2n-m}\nabla_{j]}E_k\right.  \nonumber \\
 -  & \!\!  \kappa_0  \left.\sum_{n=1}^\infty\sum_{m=1}^{2n}(-1)^{n+m}\kappa_{2n}(r)\partial_t^{m-1}B^k\partial_t^{2n-m}\nabla_{j]}B_k\right]   \nonumber \\ 
&+\varepsilon_0\sum_{n=1}^\infty\sum_{m=1}^{2n}(-1)^{n+m}\varepsilon_{2n}(r)\partial_t^{m-1}E_{[i}\partial_t^{2n-m}E_{j]}  \nonumber \\
&-\kappa_0\sum_{n=1}^\infty\sum_{m=1}^{2n}(-1)^{n+m}\kappa_{2n}(r)\partial_t^{m-1}B_{[i}\partial_t^{2n-m}B_{j]},  \label{L1} \\
M_{kij}=&\, 2\mathcal{L} r_{[i}g_{j]k}+2\varepsilon_0r_{[i}\nabla_{j]}\phi\,\varepsilon(r,i\partial_t)E_k   \nonumber \\
&+2\kappa_0\left(\epsilon_{mk[i}A_{j]}-\epsilon_{mkl}r_{[j}\nabla_{i]}A^l\right)\kappa(r,i\partial_t)B^m. \label{M1}  
\end{align}}%
The angular momentum density (\ref{L1}) and flux (\ref{M1}) satisfy the conservation law (\ref{con}) when Maxwell's Eqs.~(\ref{max1}) and (\ref{max2}) are used; one obtains
 \begin{align} 
\partial_t L^{ij}+ \nabla_k M^{kij}=&\varepsilon_0r^{[i}E_k\left[\nabla^{j]}\varepsilon(r,i\partial_t)\right]E^k   \nonumber \\
&-\kappa_0r^{[i}B_k\left[\nabla^{j]}\kappa(r,i\partial_t)\right]B^k,
\end{align}
which vanishes because $r^{[i}\nabla^{j]}f(r)=0$ for any function of $r=\sqrt{\mathbf{r}\cdot\mathbf{r}}$.

The results  (\ref{L1}) and (\ref{M1}) are not gauge invariant. Equivalent gauge-invariant expressions are found by noting that the quantities
\begin{gather}
F^{kij}=2\varepsilon_0r^{[j}A^{i]}\varepsilon(r,i\partial_t)E^k, \\
G^{lkij}=4\kappa_0r^{[j}A^{i]}\kappa(r,i\partial_t)\nabla^{[k}A^{l]}
\end{gather}
{\it identically} satisfy
\begin{equation}  \label{FGid}
\partial_t\nabla_kF^{kij}+\nabla_k(-\partial_tF^{kij}+\nabla_lG^{lkij})=0.
\end{equation}
From (\ref{FGid}) we see that adding $\nabla_kF^{kij}$ to $L^{ij}$, and adding $-\partial_tF^{kij}+\nabla_lG^{lkij}$ to $M^{kij}$, does not affect the conservation law (\ref{con}). Performing these additions, and using Maxwell's Eqs.~(\ref{max1}) and (\ref{max2}), we obtain the following gauge-invariant angular momentum density and flux:
{\allowdisplaybreaks
\begin{align}
 L_{ij} =& 2r_{[i}p_{j]} \nonumber \\
&+\varepsilon_0\sum_{n=1}^\infty\sum_{m=1}^{2n}(-1)^{n+m}\varepsilon_{2n}(r)\partial_t^{m-1}E_{[i}\partial_t^{2n-m}E_{j]}  \nonumber \\
&-\kappa_0\sum_{n=1}^\infty\sum_{m=1}^{2n}(-1)^{n+m}\kappa_{2n}(r)\partial_t^{m-1}B_{[i}\partial_t^{2n-m}B_{j]},   \label{L} \\
M_{kij}=& 2r_{[i}\sigma_{j]k},  \label{M}  
\end{align}}%
where $p_i$ and $\sigma_{ij}$ are, respectively, the linear momentum density and  stress tensor in a dispersive medium~\cite{phi11}:
{\allowdisplaybreaks
\begin{align}
 p_i=&\, (\mathbf{D}\times  \mathbf{B})_i      \nonumber \\
&+\frac{\varepsilon_0}{2}\sum_{n=1}^\infty\sum_{m=1}^{2n}(-1)^{n+m}\varepsilon_{2n}(r)\partial_t^{m-1}E_j\partial_t^{2n-m}\nabla_iE^j  \nonumber \\
&-\frac{\kappa_0}{2}\sum_{n=1}^\infty\sum_{m=1}^{2n}(-1)^{n+m}\kappa_{2n}(r)\partial_t^{m-1}B_j\partial_t^{2n-m}\nabla_iB^j,  \label{p} \\
\sigma_{ij}=&-E_iD_j-H_iB_j  +\frac{1}{2}g_{ij}\left(\mathbf{E}\cdot\mathbf{D}+\mathbf{B}\cdot\mathbf{H} \right). \label{stress}  
\end{align}}%
The linear momentum density (\ref{p}) is not the Minkowskii momentum $\mathbf{D}\times  \mathbf{B}$, which is only valid if there is no dispersion. The stress tensor (\ref{stress}) has the same expression as in the non-dispersive case, but is not symmetric in the time domain. The momentum density (\ref{p}) and stress tensor (\ref{stress}) obey a conservation law only if the medium is homogeneous~\cite{phi11}. A striking property of the angular momentum density (\ref{L}) is that it has no simple relation to the linear momentum density $\mathbf{p}$, even when linear momentum is conserved. The angular momentum flux (\ref{M}), on the other hand, is completely determined by the linear momentum flux, the stress tensor $\sigma_{ij}$. The result (\ref{M}) is the same expression for the angular momentum flux as in vacuum~\cite{jac}, except that the stress tensor is symmetric in vacuum so that $r_{[i}\sigma_{j]k}=\sigma_{k[j}r_{i]}$, whereas this last equality does not hold in a dispersive medium.

It is more usual to write optical angular momentum density as a vector, with the flux as a second-rank tensor; these quantities are simply the duals of $ L_{ij}$ and $M_{kij}$, defined by 
\begin{equation}  \label{duals}
L^i=\frac{1}{2}\epsilon^{ijk}L_{jk}, \quad M^{ij}=\frac{1}{2}\epsilon^{jkl}M^i_{\ kl}.
\end{equation}
The angular momentum density vector $\mathbf{L}$ and second-rank flux $M^{ij}$ are, from (\ref{L}), (\ref{M}) and (\ref{duals}),
{\allowdisplaybreaks
\begin{align}
\mathbf{L}=& \mathbf{r}\times\mathbf{p} \nonumber \\
&+\frac{\varepsilon_0}{2}\sum_{n=1}^\infty\sum_{m=1}^{2n}(-1)^{n+m}\varepsilon_{2n}(r)\partial_t^{m-1}\mathbf{E}\times\partial_t^{2n-m}\mathbf{E}  \nonumber \\
&-\frac{\kappa_0}{2}\sum_{n=1}^\infty\sum_{m=1}^{2n}(-1)^{n+m}\kappa_{2n}(r)\partial_t^{m-1}\mathbf{B}\times\partial_t^{2n-m}\mathbf{B},   \label{Lvec} \\
M^{ij}=& \epsilon^{jkl}r_k\sigma_{li},  \label{Mij}  
\end{align}}%
and they satisfy the conservation law
\begin{equation}
\partial_t L^{i}+ \nabla_j M^{ji}=0.
\end{equation}
The results (\ref{L}) and (\ref{M}), or their duals (\ref{Lvec}) and (\ref{Mij}), are the solution to the problem posed at the beginning of this paper.

Similar to the energy density and momentum density in a dispersive medium~\cite{phi11}, the angular momentum density simplifies considerably for time-averaged monochromatic waves. We insert the monochromatic $\mathbf{E}$ field
\begin{equation}  \label{Emono}
\mathbf{E}(\mathbf{r},t)=\frac{1}{2}\left(\mathbf{E}_0 (\mathbf{r})e^{-i\omega t}+\text{c.c}\right), 
\end{equation}
and a $\mathbf{B}$ field of the same form into (\ref{Lvec}). After time averaging, all $t$-dependent terms vanish and we obtain the time-averaged monochromatic angular momentum density $\bar{\mathbf{L}}_{\text{mono}}$:
{\allowdisplaybreaks
\begin{align}
\bar{\mathbf{L}}_{\text{mono}}=& \mathbf{r}\times\bar{\mathbf{p}}_{\text{mono}} 
-\frac{\varepsilon_0}{4}\frac{d\varepsilon(r,\omega)}{d\omega}\text{Im}(\mathbf{E}_0\times\mathbf{E}_0^*) \nonumber \\ 
&+\frac{\kappa_0}{4}\frac{d\kappa(r,\omega)}{d\omega}\text{Im}(\mathbf{B}_0\times\mathbf{B}_0^*), \label{Lmono}  
\end{align}}%
where $\bar{\mathbf{p}}_{\text{mono}}$ is the time-averaged monochromatic linear momentum density~\cite{phi11,note}
\begin{align}   
\bar{\mathbf{p}}_{\text{mono}}=& \frac{\varepsilon_0}{2}\varepsilon(r,\omega)\text{Re}(\mathbf{E}_0\times\mathbf{B}^*_0)-\frac{\varepsilon_0}{4}\frac{d\varepsilon(r,\omega)}{d\omega}\text{Im}(E_{0i}\nabla E^{i*}_0)  \nonumber \\
&+\frac{\kappa_0}{4}\frac{d\kappa(r,\omega)}{d\omega}\text{Im}(B_{0i}\nabla B^{i*}_0).  \label{pmono}
\end{align}
The time-averaged monochromatic energy density $\bar{\rho}_{\text{mono}}$ in a dispersive medium is given by the well-known Brillouin formula~\cite{jac,phi11}
\begin{equation}  \label{brill}
\bar{\rho}_{\text{mono}}=\frac{\varepsilon_0}{4}\frac{d[\omega\varepsilon(r, \omega)]}{d \omega}\mathbf{E}_0\cdot\mathbf{E}_0^*+\frac{\mu_0}{4}\frac{d[\omega\mu(r, \omega)]}{d \omega}\mathbf{H}_0\cdot\mathbf{H}_0^*.
\end{equation}
This energy result is exact only for monochromatic waves, but is often used as an approximation in the quasi-monochromatic case. The exact energy density for finite frequency ranges where absorption is negligible is given in~\cite{phi11}. Note that the energy density expression (\ref{brill}) is in fact valid for any spatial dependence of the dielectric functions, but the momentum density (\ref{pmono}) only measures a conserved quantity if the medium is homogeneous.

To study a specific example, we consider a homogeneous medium, where the dielectric functions $\varepsilon(\omega)$ and $\kappa(\omega)$ are independent of position. The simplest monochromatic wave in this case is a plane wave, but a well-known oddity of a circularly polarized plane wave in vacuum is that its angular momentum density $\varepsilon_0\mathbf{r}\times(\mathbf{E}\times\mathbf{B})$ in the propagation direction turns out to be zero~\cite{yao11}. If the circularly polarized wave is given a transverse intensity that falls off to zero at infinity, the angular momentum in the propagation direction is found to be $\pm\hbar$ per photon, the expected spin angular momentum~\cite{yao11}. To capture the full spin angular momentum of a circularly polarized monochromatic wave we must therefore consider a non-trivial transverse profile. The simplest case is to assume that the transverse profile decreases to zero slowly compared to the wavelength scale~\cite{jac}. The resulting spatial dependence $\mathbf{E}_0(\mathbf{r})$ for the monochromatic electric field (\ref{Emono}), and corresponding $\mathbf{B}_0(\mathbf{r})$, are then~\cite{jac}
\begin{gather}
\mathbf{E}_0(\mathbf{r})=\left[\mathcal{E}(y,z)(\mathbf{e}_y\pm i\mathbf{e}_z)+ \frac{i}{k}\left(\partial_y\mathcal{E}\pm i\partial_z\mathcal{E}\right)\mathbf{e}_x\right]e^{ikx}  \label{Ebeam} \\
\mathbf{B}_0(\mathbf{r})=\mp \frac{i}{c}n_p(\omega)\mathbf{E}_0(\mathbf{r}).    \label{Bbeam}
\end{gather}
We have taken the wave to propagate in the $x$-direction, with $\{\mathbf{e}_x,\mathbf{e}_y,\mathbf{e}_z\}$ unit vectors along the Cartesian axes. The transverse beam profile is given by $\mathcal{E}(y,z)$, $n_p(\omega)=\sqrt{\varepsilon(\omega)\mu(\omega)}$ is the phase index, $k=n_p(\omega)\omega/c$ is the wave vector and the upper/lower signs are for left/right circular polarization. One can verify that (\ref{Ebeam}) and (\ref{Bbeam}) satisfy Maxwell's equations when $k\mathcal{E}(y,z)$ is much larger than derivatives of $\mathcal{E}(y,z)$. With the same assumption of a slowly varying $\mathcal{E}(y,z)$, the $x$-component of the time-averaged angular momentum density (\ref{Lmono}) for the beam (\ref{Ebeam}) and (\ref{Bbeam}) is
\begin{equation} \label{Lx}
\bar{L}_x=\mp\varepsilon_0\varepsilon(\omega)\frac{n_p(\omega)}{ck}\mathcal{E}(\mathbf{r}_\perp\cdot\nabla_\perp)\mathcal{E} 
\pm\varepsilon_0\sqrt{\frac{\varepsilon(\omega)}{\mu(\omega)}}\,
\frac{dn_p(\omega)}{d\omega}\mathcal{E}^2,
\end{equation}
where $\mathbf{r}_\perp=\{y,z\}$ is the transverse position vector and $\nabla_\perp=\{\nabla_y,\nabla_z\}$ is the transverse gradient operator.
By integrating (\ref{Lx}) over the $yz$-plane we find the spin angular momentum per unit length of the beam. The first term in (\ref{Lx}) can be re-written using $2\mathcal{E}(\mathbf{r}_\perp\cdot\nabla_\perp)\mathcal{E}=\nabla_\perp\cdot(\mathbf{r}_\perp\mathcal{E}^2)-\mathcal{E}^2\nabla_\perp\cdot\mathbf{r}_\perp$ and $\nabla_\perp\cdot\mathbf{r}_\perp=2$, so the integration gives
\begin{equation} \label{Lpul}
\int d^2\mathbf{r}_\perp\bar{L}_x=\pm\frac{\varepsilon_0\varepsilon(\omega)n_g(\omega)}{\omega n_p(\omega)}\int d^2\mathbf{r}_\perp\mathcal{E}^2,
\end{equation}
where $n_g(\omega)=d[\omega n_p(\omega)]/d\omega$ is the group index. The time-averaged energy density (\ref{brill}) of the beam (\ref{Ebeam}) and (\ref{Bbeam}) is
\begin{equation}  \label{rhobeam}
\bar{\rho}=\frac{\varepsilon_0n_p(\omega)n_g(\omega)}{\mu(\omega)}\mathcal{E}^2.
\end{equation}
Integration of (\ref{rhobeam}) over the $yz$-plane gives the energy per unit length, and dividing this into (\ref{Lpul}) we find a spin angular momentum per unit energy of $\pm\omega^{-1}$, just as in vacuum~\cite{jac}. An energy of $\hbar\omega$ in the beam thus carries a spin angular momentum of $\pm\hbar$.

The time-averaged momentum density (\ref{pmono}) of the beam (\ref{Ebeam}) and (\ref{Bbeam}) in the $x$-direction is
\begin{equation} 
\bar{p}_x=\frac{1}{c}\varepsilon_0\varepsilon(\omega)n_g(\omega)\mathcal{E}^2,
\end{equation}
and dividing this by the time-averaged energy density (\ref{rhobeam}) gives a momentum per unit energy of $n_p(\omega)/c$. An energy of $\hbar\omega$ in the beam therefore carries a linear momentum of $\hbar k$ (this is also the result for a plane wave).

Although we have calculated spin angular momentum and linear momentum for an energy $\hbar\omega$, the results presented here are purely classical. Nevertheless, we have shown a highly non-trivial interplay of dispersive contributions to the classical energy, momentum and angular momentum of light that indicates the expected results for photon momentum and angular momentum. A rigorous quantization of light in dispersive media requires a full account of absorption consistent with Kramer-Kronig relations~\cite{phi10}. The energy-momentum tensor of classical and quantum light in an arbitrary medium obeying Kramer-Kronig relations is given in~\cite{phi11b} in a form that includes the energy-momentum absorbed by the medium. The total angular momentum density and flux in an medium obeying Kramer-Kronig relations can easily be found from the action in~\cite{phi10,phi11b} using Noether's theorem. Separation of total energy-momentum and angular momentum into electromagnetic and absorbed parts is a difficult problem in general, but for frequencies with negligible absorption the results derived here and in~\cite{phi11} must follow from a more general treatment that includes losses.

As we stressed in the opening paragraphs, we have not analyzed angular momentum transfer from light to matter. This transfer can occur either by absorption of a pulse carrying angular momentum or by transmission of the pulse through the medium. The results derived here can be used to calculate the total angular momentum carried by the pulse when it is completely contained inside the medium. An analysis of the angular momentum transfered to the material~\cite{pad03} leads to the same subtleties that have been long debated in the case of linear momentum (the Abraham-Minkowskii controversy~\cite{bre79,pfe07}).

This research was supported by the Royal Society of Edinburgh and the Scottish Government.


\begin{thebibliography}{99}
\bibitem{all92}
L.\ Allen, M.\ W.\ Beijersbergen. R.\ J.\ C.\ Spreeuw, and J.\ P.\ Woerdman, Phys.\ Rev.\ A {\bf 45}, 8185 (1992). 

\bibitem{yao11}
A.\ M.\ Yao and M.\ J.\ Padgett, Adv.\ Opt.\ Photon.\ {\bf 3}, 161 (2011). 

\bibitem{jac}
J.\ D.\ Jackson, {\it Classical Electrodynamics,} 3rd ed.\ (Wiley, New York, 1999).

\bibitem{phi11}
T.\ G.\ Philbin, Phys.\ Rev.\ A {\bf 83}, 013823 (2011). Erratum: Phys.\ Rev.\ A {\bf 85}, 059902(E) (2012).

\bibitem{pfe07}
R.\ N.\ C.\ Pfeifer, T.\ A.\ Nieminen, N.\ R.\ Heckenberg, and H.\ Rubinsztein-Dunlop, Rev.\ Mod.\ Phys.\ {\bf 79}, 1197 (2007). 

\bibitem{note}
Note that the Erratum in~\cite{phi11} corrects an error in the monochromatic momentum result.

\bibitem{pad03}
M.\ J.\ Padgett, S.\ M.\ Barnett, and R.\ Loudon, J.\ Mod.\ Opt.\ {\bf 50}, 1555 (2003). 

\bibitem{man08}
M.\ Mansuripur, Opt.\ Express {\bf 16}, 14821 (2008). 

\bibitem{agr}
G.\ Agrawal, {\it Nonlinear Fiber Optics,} 4th ed. (Academic Press, San Diego, 2006).

\bibitem{wei}
S.\ Weinberg, {\it The Quantum Theory of Fields,} Volume~I (Cambridge University Press, Cambridge, 1995).

\bibitem{phi10}
T.\ G.\ Philbin, New J.\ Phys.\ {\bf 12}, 123008 (2010). 

\bibitem{phi11b}
T.\ G.\ Philbin, New J.\ Phys.\ {\bf 13}, 063026 (2011). 

\bibitem{bre79}
I.\ Brevik, Phys.\ Rep.\ {\bf 52}, 133 (1979).

\end{thebibliography}
\end{document}